# Fractal signature as a rotational modulation and stellar noise classifier based on the active Kepler stars

# Assinatura fractal como um classificador de ruído estelar e modulação rotacional baseada nas estrelas ativas da Missão Kepler

Paulo Cleber Farias da Silva Filho
M.A. student in Astrophysics
Institution: Federal University of Ceará
Address: Campus do Pici, Fortaleza – CE
E-mail: clebersilva@fisica.ufc.br

José Ribamar Dantas Silveira Júnior
Graduate student in Astrophysics
Institution: Federal University of Ceará
Address: Campus do Pici, Fortaleza – CE
E-mail: ribamar@fisica.ufc.br

Brício Warney de Freitas Alves
M.A. student in Astrophysics
Institution: Federal University of Ceará
Address: Campus do Pici, Fortaleza – CE
E-mail: bricio@fisica.ufc.br

Fernando José Silva Lima Filho
Undergraduate student in Physics
Institution: Federal University of Ceará
Address: Campus do Pici, Fortaleza – CE
E-mail: fernandolima@fisica.ufc.br

Vitor Marcelo Belo Ferreira
Undergraduate student in Physics
Institution: Federal University of Ceará
Address: Campus do Pici, Fortaleza – CE
E-mail: vitormarcelobeloferreira@alu.ufc.br

Luiz Daniel Alves Rios
PhD student in Astrophysics
Institution: Federal University of Ceará
Address: Campus do Pici, Fortaleza – CE

E-mail: daniel_alvesrios@hotmail.com

Thiago de Melo Santiago
PhD student in Astrophysics
Institution: Federal University of Ceará
Address: Campus do Pici, Fortaleza – CE
E-mail: thiago@fisica.ufc.br

Daniel Brito de Freitas
Professor Doctor
Institution: Federal University of Ceará
Address: Campus do Pici, Fortaleza – CE
E-mail: danielbrito@fisica.ufc.br

**ABSTRACT**

In this study, we report on the analysis of 701 stars in a solar vicinity defined in three categories namely subsolar, solar, and supersolar with rotation periods between 1 and 70 days, based on rotational modulation signatures inferred from time series from the Kepler mission's Public Archives. In our analysis, we performed an initial selection based on the rotation period and position in the period–$H$ diagram, where $H$ denotes the Hurst exponent extracted from fractal analysis. To refine our analysis, we applied a fractal approach known as the R/S method, taking into account the fluctuations of the features associated with photometric modulation at different time intervals and the fractality traces that are present in the time series of our sample. In this sense, we computed the so-called Hurst exponent for the referred stars and found that it can provide a strong discriminant of rotational modulation and background noise behavior, going beyond what can be achieved with solely the rotation period itself. Furthermore, our results emphasize that the rotation period of stars is scaled by the exponent $H$ which increases following the increase in the rotation period. Finally, our approach suggests that the referred exponent may be a powerful rotational modulation and noise classifier.

**RESUMO**

Neste estudo, relatamos a análise de 701 estrelas em uma vizinhança solar definida em três categorias: subsolar, solar e supersolar com períodos de rotação entre 1 e 70 dias, com base em assinaturas de modulação rotacional inferidas de séries temporais dos arquivos públicos da missão Kepler. Em nossa análise, realizamos uma seleção inicial baseada no período de rotação e posição no diagrama período–*H*, onde *H* denota o expoente de Hurst extraído da análise fractal. Para refinar nossa análise, aplicamos uma abordagem fractal conhecida como método R/S, levando em consideração as flutuações das características associadas à modulação fotométrica em diferentes intervalos de tempo e os traços de fractalidade presentes nas séries temporais de nossa amostra. Nesse sentido, calculamos o chamado expoente de Hurst para as referidas estrelas e descobrimos que ele pode fornecer um forte discriminante da modulação rotacional e do comportamento do ruído de fundo, indo além do que pode ser alcançado apenas com o próprio período de rotação. Além disso, nossos resultados enfatizam que o período de rotação das estrelas é dimensionado pelo expoente *H* que aumenta com o aumento do período de rotação. Por fim, nossa abordagem sugere que o referido expoente pode ser um poderoso classificador de modulação rotacional e ruído.

**Palavras-Chave:** ruído estelar, missão Kepler (NASA), modulação rotacional, análise fractal

## 1 INTRODUCTION

Many physical phenomena, such as these found in astrophysical time series, seem to be controlled either by rare or common events, which is the case of flares and starspot, respectively. These events reveal strong (multi)fractal properties. As an example, de Freitas et al. (2013) discussed the statistical and fractal properties of CoRoT time series due to cyclic behavior attributed to the magnetic activities on the stellar photosphere. The importance of the astrophysical time series analysis, which exhibits typically nonlinear dynamics, has been recognized in complex systems analysis. Several features of these approaches have been adopted to detect the time-dynamical behavior of the astrophysical phenomena (cf. Mandelbrot & Wallis, 1969).

The variations of Hurst exponent $H$ in different time windows can quantify the dynamic changes of the features of time series. As an example, lower $H$ values characterize the portions of the signal with lower complexity. In contrast, upper values reveal strong interactions in dynamics of time series. This behavior clearly violates the Boltzmann-Gibbs (B-G) statistics. Likewise, different levels of complexity can be associated to anti-persistency/persistency (fractional Brownian motion or not) processes extracted by a fractal spectral analysis in terms of the global or local Hurst exponent. As mentioned by de Freitas et al. (2013), the exponent $H$ can indicate a quantification of variability in relatively brief and noisy time series.

In this sense, nonlinear evolving dynamical systems are a class of physical ensembles involving long-range interactions (complex fluctuations), long-range microscopic memories (e.g., non markovian stochastic processes), or (multi)fractal structures. In this context, time-domain astrophysics can hardly be treated within the traditional statistical analysis, particularly Fourier analysis, for a wide range of problems (Feigelson & Babu, 2012). In general, structural properties in astrophysical time series are clearly achieved without using robust methods. Therefore, several sources of (multi)fractality are not captured by conventional measures like the Fourier and Wavelet transforms of the signal. A signal like the rotational modulation or magnetic activity in the small structures on the stellar photosphere possess a scale invariant structure. For instance, a rotational signature has a scale invariant structure when a given structure repeats itself on regular subintervals of the time. Mathematically, the astrophysical time series $X(t)$ are scale invariant when $X(ct) = c^H X(t)$, where $c$ is a constant (Ihlen, 2012). This exponent defines a kind of global scale invariant present in the signal. Signals that exhibit a structure independent on time and space are known as monofractals and are defined by a single power law exponent, i.e., one unique value for $H$. However, spatial, and temporal variations in scale invariant structure are most common in astrophysical scenario. In this case, time series are best defined by a with spectrum of $H$ (cf. de Freitas et al., 2013; Filho et al. 2022).

Some authors (e.g., Suyal et al., 2009 and de Freitas et al., 2013) have shown that the variations of stellar magnetic activity present a global value of $H$ great than 0.5 indicating a long-term memory in the time series. In particular, the Hurst exponent is used as a measure of long-term memory of time series (Kantelhardt et al., 2002). However, a multifractal time series not only exhibit long-range correlations, but also result from fat tails in probability distributions even if there is no memory in the data (Grech & Pamula, 2008). On finite samples of data, large fluctuations cannot be detected, contrary to small fluctuations. In other words, the multifractal properties in shorter time series reveal a wide variety of autocorrelations at different scales, while longer time series are corrupted by various effects like noise, short-term memory, or periodicity in signal (Grech & Pamuła, 2008).

de Freitas et al. (2013) found the fractality traces present in the time series of the Sun in its active and quiet phases. They also found them on a sample of 14 CoRoT stars with periods of subsolar and super solar rotations (less than 24 days and greater than 34 days, respectively) and 3 stars with periods within the solar values (between 24 and 34 days). They computed the global Hurst exponent for these stars and found a strong correlation between the Hurst exponent and the rotational period. As a result, these authors showed that the Global Hurst exponent can be indicate as a powerful classifier for semi-sinusoidal time series. However, this method should give different results for the exponent *H* in the astrophysical noise level. In other words, changes in the Hurst exponent can indicate a correlation between the noise and the original signal.

The main aim of the present work is to examine the dependence of global exponent *H* with the stellar variability in short time series less than 150 days, a phenomenon that has been observed in other works as in de Freitas et al. (2013). Indeed, we will investigate the behavior of variability of Kepler mission's time series with rotation period from few days to ~70 days. In this context, we will compare the values of the Hurst exponents for original, differencing filter, shuffled and randomized data and to verify if the processes are anti-persistent (anti-correlated) or mean-reverting, among other features.

The present paper is organized as follows. In Section 2 we describe the data used in our study. In Section 3, we describe the methods used in our analysis of these data. In Section 4, we provide our results and discuss their implications. Finally, concluding remarks are presented in last section.

**2 WORKING SAMPLE**

From 2009 May to 2013 May the *Kepler* mission performed 17 observational runs of ~90 days each (hereafter quarters), composed of long cadence (data sampling every 29.4 min, Jenkins et al., 2010b) and short cadence (sampling every 59 s) observations (Van Cleve et al., 2010; Thompson et al., 2013). Detailed discussion of public archive can be found in many *Kepler* team publications, *e.g.*, Borucki et al. (2009, 2010), Batalha et al. (2010), Koch et al. (2010), and Basri et al. (2011). Regarding data format, *Kepler* archive provides both Simple Aperture Photometry data (processed using a standard treatment only removing more relevant spacecraft artifacts) and Pre-Search Data Conditioning (PDC) data, processed through a refined treatment based on the *Kepler* pipeline (Jenkins et al. 2010a) which removes more of thermal and kinematics effects due to spacecraft operation (Van Cleve & Caldwell, 2009).

**Figure 1:** An extract of three stars from our sample. Time series with temporal window of 120 days for the Kepler stars: KIC 002713086 (top panel), KIC 001027740 (middle panel) and KIC 006605595 (bottom panel) with rotation periods of 4.14, 26.7 and 51.9 days, respectively.

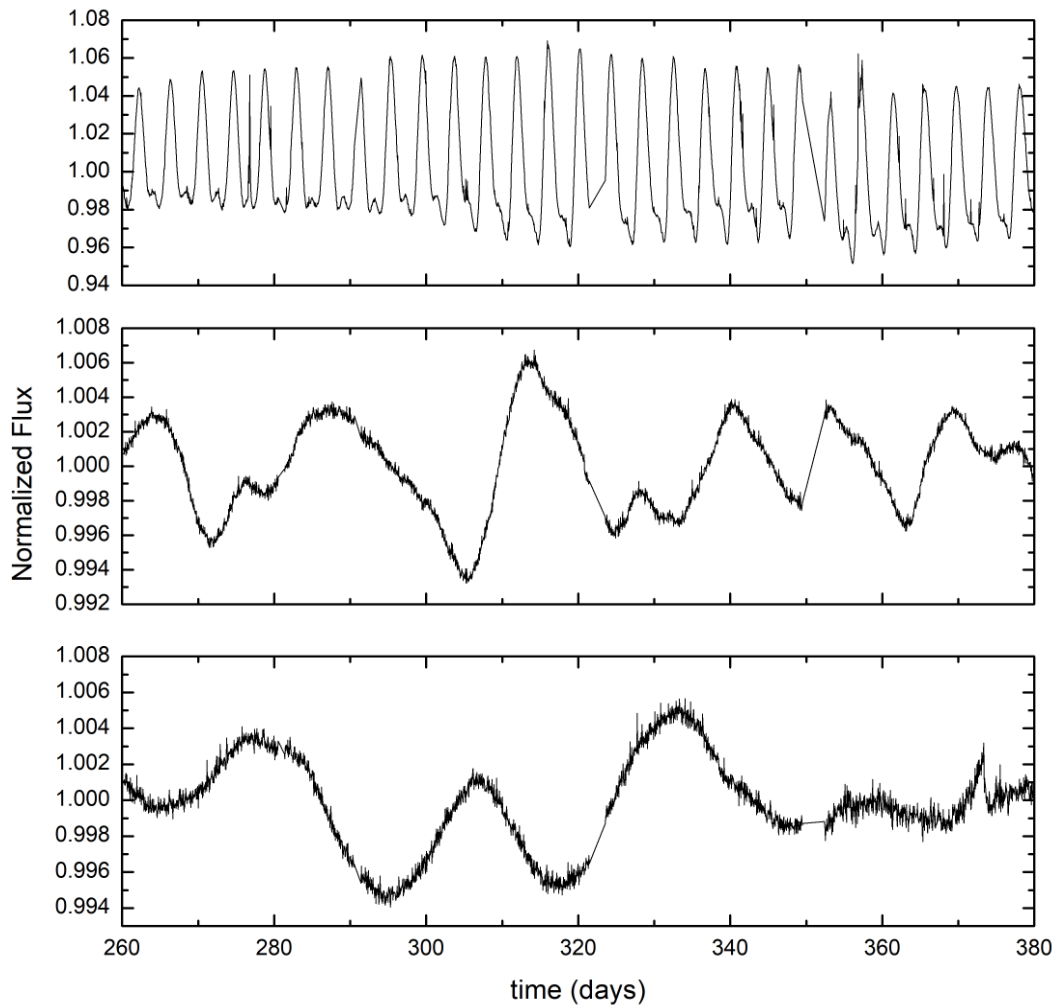

Quarter to quarter offsets were treated as in Banyai et al. (2013) and De Medeiros et al. (2013), and outliers out of 3.5 times standard deviation from a 3rd order polynomial fit to each quarter were removed. As PDC light curves are in general good quality time series, only few manual treatments were needed, removing artifacts and instrumental trends (see Petigura & Marcy, 2012) of only 13 stars from the complete sample. Additionally, visual inspection of the entire sample gets 20 stars to be removed from the solar sample due to higher systematic effects. In the construction of our light curves (hereafter LC), only quarters 1 to 16 were considered in its PDC pipeline. For the sake of our analysis, cadence was set two times the original long cadence and thus sampling became 59 min.

Our working sample consists of 644 stars in a solar vicinity defined by $T_{\text{eff}}$ between 5579 and 5979 K, $\log g$ between 3.94 and 4.94 dex, and $P_{\text{rot}}$ between 24 and 34 days (see Figure 1). There were also defined two comparison samples named sub- and supersolar, using the same criteria for $T_{\text{eff}}$ and $\log g$ ranges but with rotational periods outside the previous range. Those samples were obtained from Chaplin et al. (2011) raising 17 subsolar objects, and from McQuillan et al. (2013) raising 13 subsolar jointly with 27 supersolar objects. Thus, our final sample consists of 701 stellar targets. The period interval for the above-mentioned selection was based on the results from Lanza et al. (2003), on which the limits for Sun rotational periods was set 24.5 days (equator) and 33.5 days (poles). Values of rotational periods estimated through auto-correlation function were taken from McQuillan et al. (2013), while temperature and gravity were from Pinsonneault et al. (2012) (SDSS corrected temperature and KIC surface gravity).

Even when McQuillan et al. (2013) periods estimate shows a high confidence for most of the nowadays conducted analysis, we highlight the present work relies in a deeper framework, which would be more affected by "lower" uncertainties. In fact, the (multi)fractal analysis (as to be explained on next section) does not reflects only an overall behavior but a detailed pattern of how different variability levels interact and dominate or shade. Thus, we defined a compromise between the required sample size and the accuracy of the rotational measurements. This balance led us to the present sample.

**3 HURST EFFECT ON THE ORIGINAL AND SURROGATE TIME SERIES**

In their work, de Freitas et al. (2013) used the well-known rescaled range (*R/S*) method proposed by Mandelbrot & Wallis (1969b) for obtaining the global Hurst exponent (hereafter *H*). The method consists of finding the relationship between the range of partial sums of deviations of sequences from its mean *R*, rescaled by its standard deviation $S$ as a function of timescale. Then, the *R/S* method is estimated by computing the sub-sample mean $\bar{y}_s = \frac{1}{M}\sum_{k=1}^{M} y_k$, where $M = sN$, $s$ is limited between 0 and 1, and $N$ denotes the length of time series. Thus, the relationship between $R_s = \max\{z_i\} - \min\{z_i\}$, where $z_i = \sum_{k=1}^{i}(y_k - \bar{y}_s)$, and the sample standard deviation $\sigma_s = \left[\frac{1}{M}\sum_{k=1}^{M}(y_k - \bar{y}_s)^2\right]^{1/2}$ defines the rescaled range as $\left(\frac{R}{S}\right)_s = \frac{R_s}{\sigma_s}$. If the process is stochastic, the *R/S* statistics follow a power-law, $\left(\frac{R}{S}\right)_s = Cs^H$, that defines the global Hurst exponent as:

$$H = \frac{\log\left(\frac{R}{S}\right)_s - \log C}{\log s} \qquad \text{(eq. 1),}$$

where $C$ is a constant.

As argued by Mandelbrot (1983), the *R/S* method is a powerful tool for detecting long-term memory and fractality of a time series, when compared to more conventional approaches, such as autocorrelation analysis and variance ratios. Differently from de Freitas et al.'s work, we used an alternative version of the *R/S* method proposed by Cannon et al. (1997). In this version known as scaled windowed variance, we estimate $H$ of a fractional Brownian motion time series by linear regression as shown in eq. 1. In addition, a time series is repeatedly divided into 10 windows, thus the means of standard deviations in such window are used to calculate an estimate of $H$. For further details about the *R/S* method, the reader can refer to the de Freitas et al. (2013).

Different values of $H$ imply fundamentally different variability behaviors on a time series. Values of $H$ equal to 0.5 show that a time series is an independent and identically distributed (i.i.d.) stochastic process, i.e., a purely Brownian motion. For values between 0 and 0.5 a time series is anti-persistent, that is, the variability follows a mean reverting process. Finally, if $H$ is between 0.5 and 1, a time series is considered persistent with long-term memory. Broadly speaking, in a time series, if the dynamics that governs the variability is not known or if the signal is noisy, it is important to investigate the different sources of small and large fluctuations.

In this context, as cited by Strozzi et al. (2007), the method of surrogate data consists in a robust procedure to understand if the variability presents in the time series is most likely due to nonlinearities or due to random inputs (Theiler et al., 1992). Likewise, this procedure can be used in combination with differencing filter. For a given time series $y(1), y(2), \ldots, y(N)$, this filter consists in difference between the next-neighbor points, such as $y(i + 1) - y(i)$. According to Feigelson (2012, p. 294, 397), the differencing filter is commonly used to remove trends of various types (global, periodic, or stochastic) to reveal short-term structure. However, it preserves the anticorrelation features.

**Figure 2:** Cumulative distribution of exponent $H$ of original data divided in three samples (subsolar, solar and supersolar).

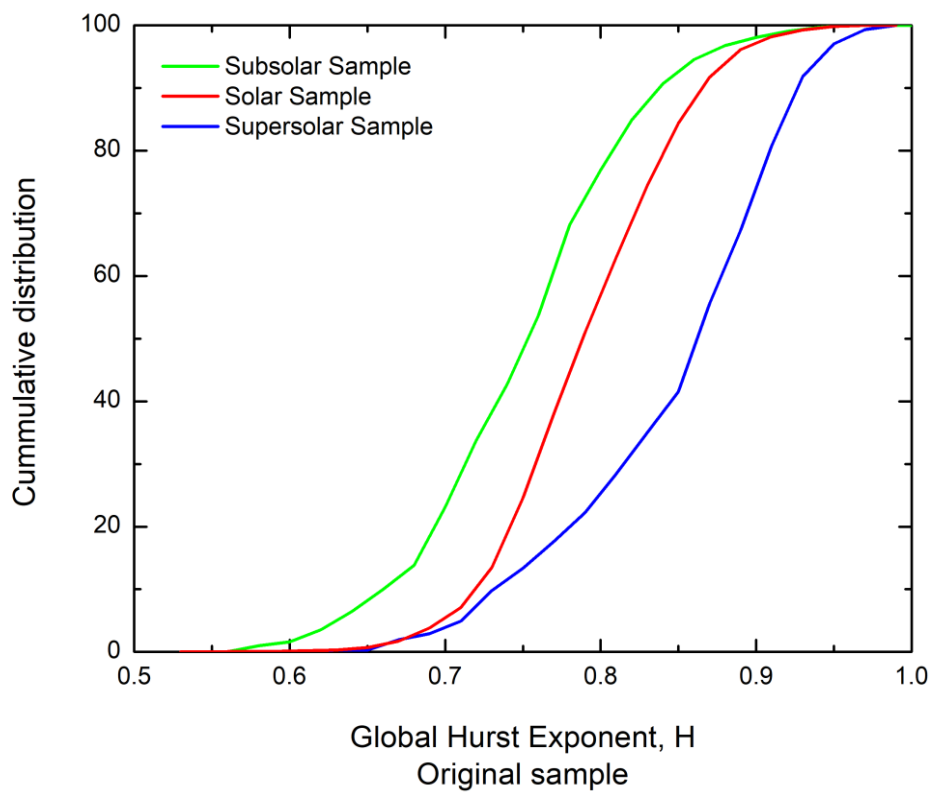

In other two procedures, we investigate the effects due to different long-time correlations for weak and strong fluctuations and due to heavy tails in probability distributions. Respectively, these two ways are related to shuffling and phase randomization of the data. The shuffling of time series destroys the long-range correlation and preserves the distribution of the fluctuations. In other words, the fluctuation distribution remains the same, but without memory. For phase randomization, the amplitudes of the Fourier transform are preserved, but the Fourier phases are randomized (Norouzzadeha et al., 2007).

**4 RESULTS AND DISCUSSIONS**

As observed in Figure 2, the histogram shows a clear difference among the means (central values) in such distribution. By visual inspection, the profile of the histogram reveals that the populations are sharply dissimilar. This difference was quantified using several tests, such as Komolgorov-Smirnov and Anderson-Darling, obtaining probabilities $P < 10^{-4}$. This values strongly indicate that the three distributions are drawn from different parent distributions.

Based on the values of $H$ computed in the present study, we can tentatively define an analytical relation between the global Hurst exponent and the period as suggested by de Freitas et al. (2013). According to these authors, the Hurst exponent may indeed be a powerful new classifier for semi-sinusoidal light curves, and its behavior can be predicted by relationship:

$$H = aP_{\text{rot}}^b + c \qquad \text{(eq. 2)},$$

where $P_{\text{rot}}$ is the rotation period and the letters $a$, $b$ and $c$ are constant (cf. de Freitas et al., 2017). The tendency of data points plotted in Figure 3 can be nicely described by the above expression.

As we can see in the Figure 3, there is a wide dispersion mainly around the solar values. To reduce these effects, we apply various filters, as well as differentiate the time series. For the latter, the results were not satisfactory, since the procedure was not able to reduce the long-term effects that can generate this behavior.

In addition, there are other types of filters that can be useful in time series analysis (cf. Woodward, Gray and Elliot, 2017, p. 66-67). Filters such as low pass and high pass can be applied in the time domain before spectral analysis of the original signal. A low pass filter (LPF) allows components of the original signal with frequencies below a certain value to be maintained after convolution, while a high pass filter allows values above this threshold to be maintained. Both works nulling the Fourier coefficients for a determined range of frequencies (cf. Samani, 2019, p. 71).

**Figure 3:** Average values of the Hurst exponents *H* calculated at the ten-time windows as a function of rotation period, for all stars in our sample. This exponent was derived based on our *R/S* analysis. The gray solid curve denotes the analytical expression given by eq. 2 that best fits the data using the same values obtained by de Freitas et al. (2013).

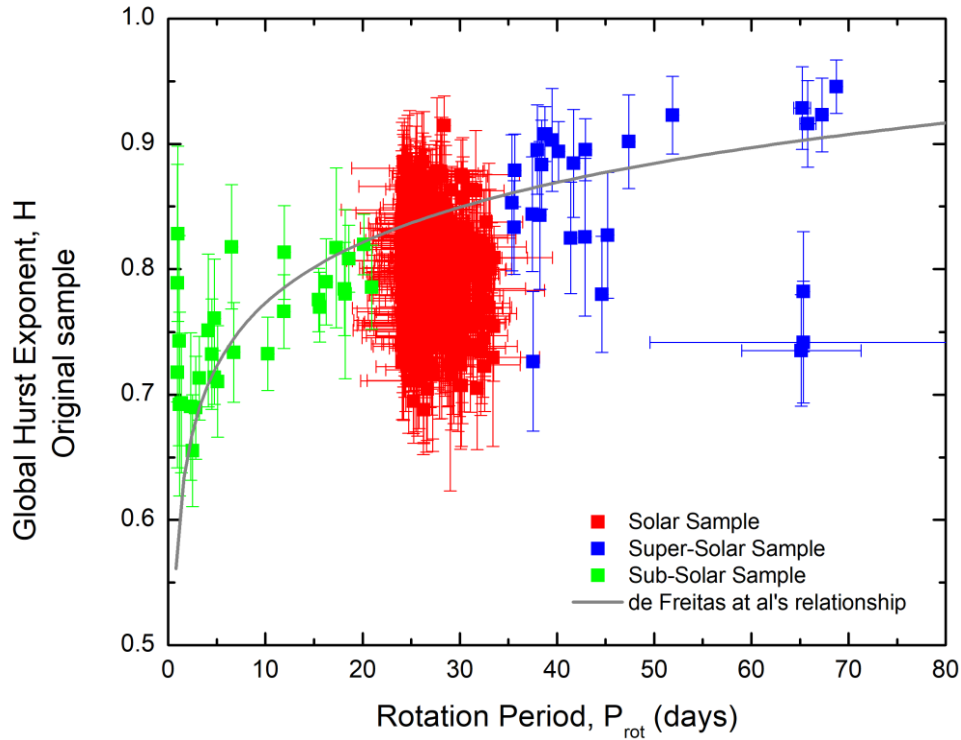

A LPF can be used to reduce variance from short-term variations and estimate the monotonic trend as a high-pass filter can remove aperiodic long-term trends (cf. Feigelson and Babu, 2012, p. 294, 397). These filters can be used to convolution of original time series, producing filtered signals that can be investigated in the linear approach. By using the LPF, we verified that the procedure significantly affects fast rotators, raising, on average, the values of *H* as can be seen in Figure 4 (green squares). This is mainly because the filter attenuates signals with frequencies higher than the cutoff frequency.

**Figure 4:** Average values of the Hurst exponents *H* as a function of rotation period, for our smoothed original sample using LPF. This exponent was derived based on our *R/S* analysis. The gray solid curve denotes the analytical expression given by eq. 2 that best fits the data using the same values obtained by de Freitas et al. (2013).

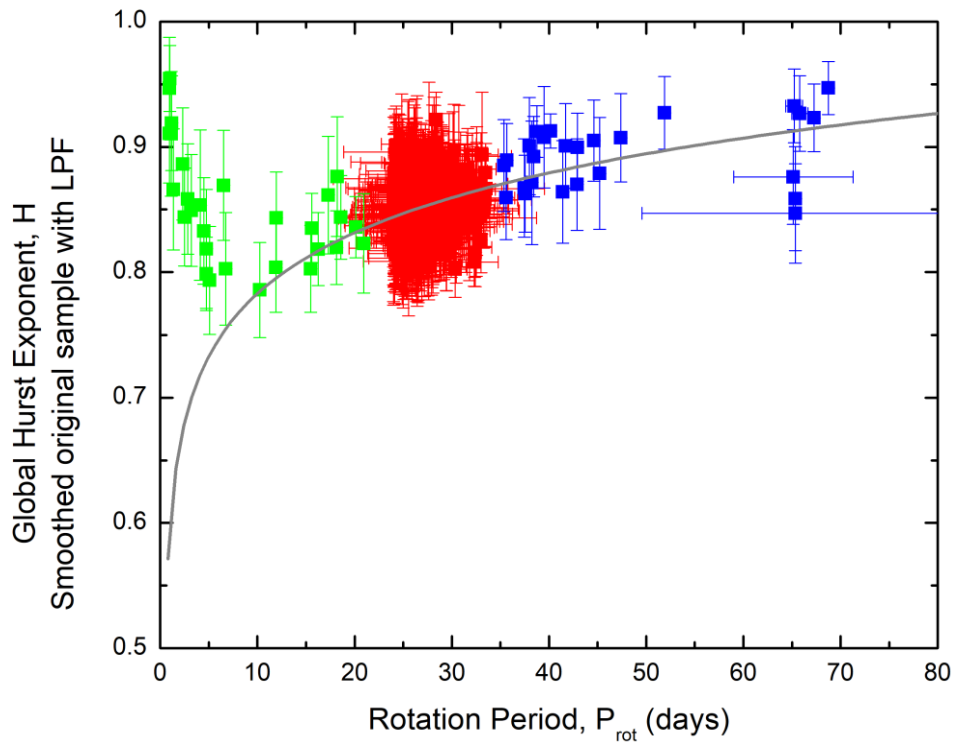

Based on these results, we decided to find another way to investigate the relationship between the *H* exponent and the rotation period. In this sense, we select only stars that have solar-like density. The motivation for such an attitude comes from the fact that there may be a bias in the sample associated not only with the mass but also with the stellar radius. With this procedure, we hope to reduce the dispersion and visualize the referenced relationship more clearly. As shown in Figure 5, the relationship seems evident and with signs that the increase in *H* follows the increase in the rotation period. However, the scattering around the solar value was not completely reduced, but it shows that the highest density of points is within the error bar, as indicated in the Figure 5. Our best fits point out that the rotation period-*H* relationship is given by

$$\log H = (-0.193 \pm 0.03) + (0.1 \pm 0.01) \log P_{\text{rot}} \qquad \text{(eq. 3)}.$$

A relationship like that was obtained by de Freitas et al (2019a). The authors suggest that the Hurst exponent is a powerful classifier of astrophysical time series when they present noise and rotational modulation.

**Figure 5:** Log-log plot of Hurst exponents *H* as a function of rotation period, considering only stars from solar sample with solar density $\rho = 1,41 \pm 0.1$ g/cm$^3$. The gray solid curve denotes

the analytical expression given by eq. 2 that best fits the data using the same values adopted by de Freitas et al. (2013).

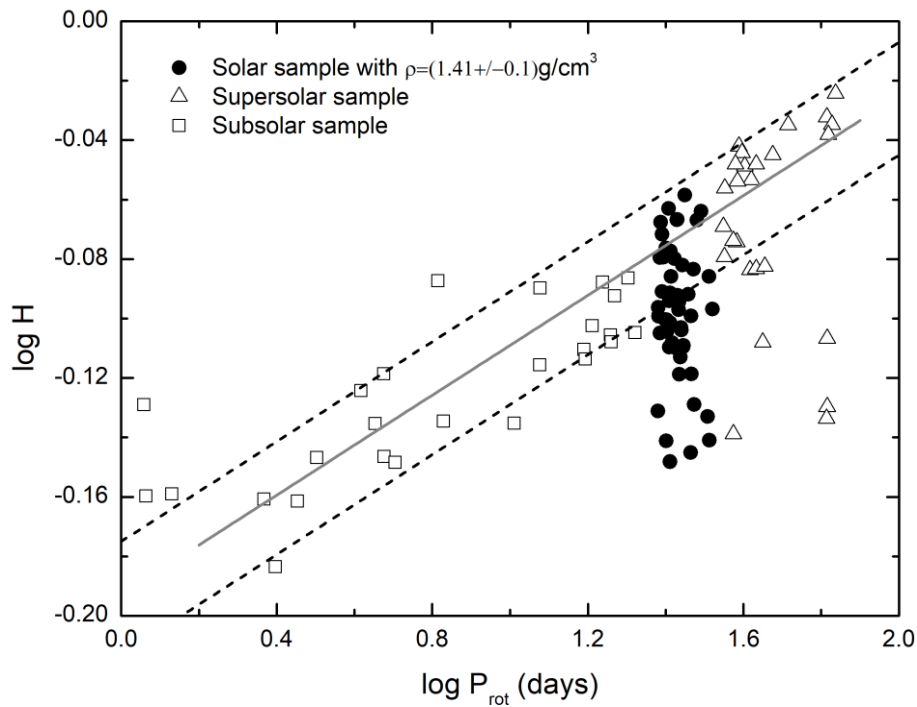

As mentioned in the previous section, there are two possible sources for explaining the origin of fractality. First, the long-term correlations due to memory and, second, the fat-tailed probability distribution due to nonlinearity. In the case of Kepler time series here analyzed, it is shown that the first source of fractality applies and that the second one has no effect (de Freitas et al., 2016, 2017, 2019a, 2019b, 2021). Figures 6 and 7 highlight the values of Hurst exponent calculated from the average results of 200 realizations of the shuffled and phase-randomized surrogates. According to Figure 6, the values of *H* of the phase-randomized series are similar to the ones of the original series. In fact, as can be seen in Fig. 7, the shuffled procedure destroyed the correlations, i.e., the values of *H* are slightly flat ($H\sim 0.5$). Thus, this procedure reduces the time series to a white noise-like one. These findings suggest that the fractality of rotational modulation is due only to long-range correlation.

**Figure 6:** Correlation between the values of *H* obtained by using original sample and phase-randomization time series. According to high correlation, the source of fractality is not due the fat-tailed probability distribution.

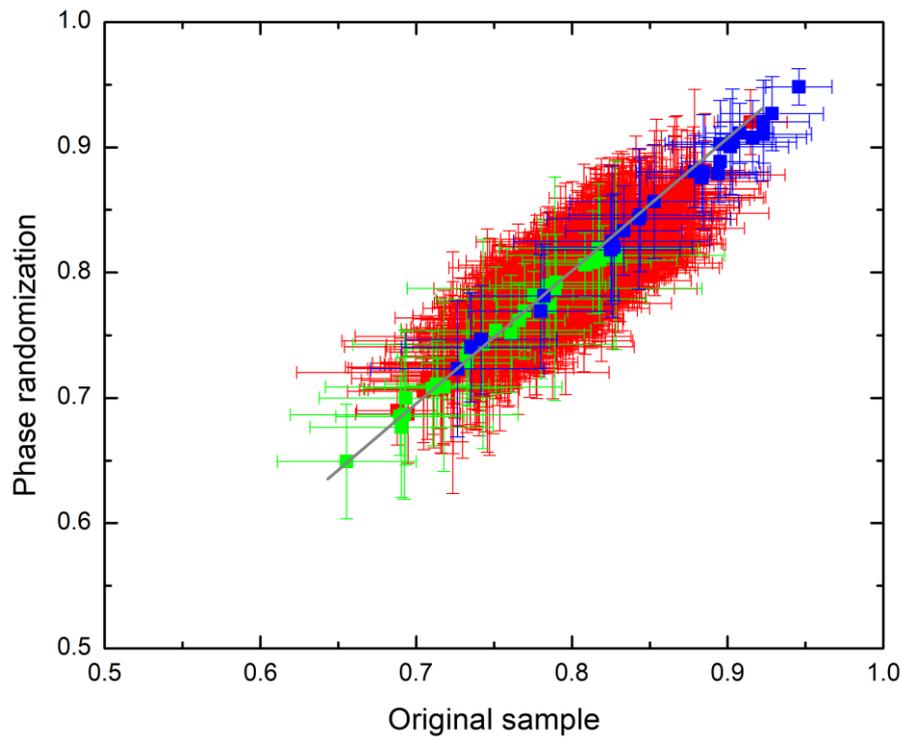

This analysis of the fractal properties reveals that the H index is strongly linked to the rotational period as proposed by de Freitas et al. (2013). Consequently, the stellar rotation is associated with the degree of persistence of the signal because of the statistical properties of H. It is worth noting that this result strengthens the general proposition that the H index is related to the evolution of stellar angular momentum and magnetic activity, as well as properties associated with the magnetic cycle. Finally, this scenario provides the theoretical basis to support the idea that the exponent is a powerful rotational modulation and noise classifier.

**Figure 7:** Correlation between the values of *H* obtained by using original sample and shuffled time series. As the values of *H* were reduced to approximately 0.5, the source of fractality is due the long-term correlation because the rotational modulation.

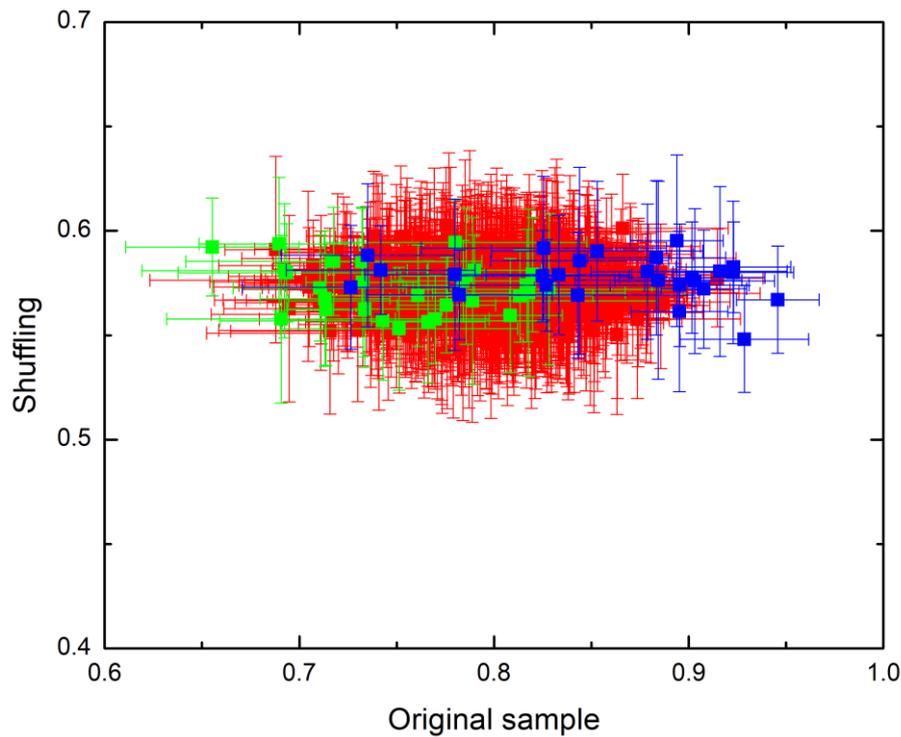

## 5 FINAL REMARKS

The present work reports three stellar samples with rotation features solar, subsolar, and supersolar. The investigated stars have rotation periods ranging from 1 to 70 days, whereas the Sun's rotation ranges from 23 days at the equator to 33.5 days at the poles. Our work, based on public data obtained by the Kepler satellite, also offers a new and exciting procedure for the treatment of stellar light curves, once we apply a powerful approach based on complex systems. More particularly, on fractality properties associated with the signal using the well-known *R/S* method. This procedure was already applied to the CoRoT data, but to our knowledge, this is the first time that it is applied to stellar light curves observed by the Kepler mission. We are confident that the proposed approach will be also important for the treatment of rotation, activity, and related phenomena in other stars of the TESS and JWST missions, as well as for stars of the future PLATO mission. One important aspect of this new procedure is the fact that we can identify, based on the Hurst exponent behavior, the level of stellar activity and related phenomena.

We used the Hurst effect to investigate the fractal signature of 701 time series as a powerful classifier of stellar noise and rotational modulation. From that analysis, we found that there is a perceptible relationship between the Hurst exponent $H$ and the rotation period $P_{\text{rot}}$, as illustrated by Figs. 3, 4 and 5.

Our results reveal a strong long-range persistence in the time series studied, indicated by the values of $H$ greater than 0.5. The long-range persistence shows that the coupling between random variables in astrophysical time series at different times is stronger than the short-range one, that is, the stellar noise. In other words, the persistent behavior in the time series characterized by calculating the Hurst exponent shows that the existence of the persistence of long-range memory as can be observed by Fig. 7. In addition, there is a clear-cut correlation between the rotation period from three stellar samples and $H$ when they are segregated by density, as shown in Fig. 5.

The existence of an empirical correlation between the Hurst exponent and the rotation period opens up the possibility of proposing a better model for explaining a variety of behaviors in the magnetic activity of the stars studied here. Given the clear difference between the values of $H$ found for the time series of stars in the original and surrogate categories, stellar noise and rotation are driven by different physical mechanisms, even if part of the noise has a magnetic origin. Furthermore, these results indicate that stellar magnetic activity behavior is even more complex than models can extract using classical models, such as LPF and Fourier transform. Finally, this paper poses a great observational and theoretical effort to go forward with simple models of stellar rotation and stellar noise.

**Acknowledgments**

We acknowledge financial support from the Brazilian agency CNPq-PQ2 (Grant No. 305566/2021-0). Research activities of STELLAR TEAM of Federal University of Ceará are supported by continuous grants from the Brazilian agency CNPq. This paper includes data collected by the *Kepler* mission. Funding for the *Kepler* mission is provided by the NASA Science Mission directorate. All data presented in this paper were obtained from the Mikulski Archive for Space Telescopes (MAST).